\documentclass[preprint, prb, showpacs, preprintnumbers,amsmath,amssymb]{revtex4-1}

\usepackage{graphicx}
\usepackage{dcolumn}
\usepackage{bm}

\newcommand{\beq}{\begin{equation}}
\newcommand{\eeq}{\end{equation}}
\newcommand{\bdis}{\begin{displaymath}}
\newcommand{\edis}{\end{displaymath}}
\newcommand{\bea}{\begin{eqnarray}}
\newcommand{\eea}{\end{eqnarray}}
\newcommand{\barr}{\begin{array}}
\newcommand{\earr}{\end{array}}
\newcommand{\bfig}{\begin{figure}[!]}
\newcommand{\efig}{\end{figure}}

\begin{document}

\preprint{}

\title{Non-equilibrium thermodynamics of the longitudinal spin Seebeck effect}

\author{Vittorio Basso, Elena Ferraro, Alessandro Magni, Alessandro Sola, Michaela Kuepferling and Massimo Pasquale}
\affiliation{Istituto Nazionale di Ricerca Metrologica, Strada delle Cacce 91, 10135 Torino, Italy}

\date{\today}


\begin{abstract}
In this paper we employ non equilibrium thermodynamics of fluxes and forces to describe magnetization and heat transport. By the theory we are able to identify the thermodynamic driving force of the magnetization current as the gradient of the effective field $\nabla H^*$. This definition permits to define the spin Seebeck coefficient $\epsilon_M$ which relates $\nabla H^*$ and the temperature gradient $\nabla T$. By applying the theory to the geometry of the longitudinal spin Seebeck effect we are able to obtain the optimal conditions for generating large magnetization currents. Furthermore, by using the results of recent experiments, we obtain an order of magnitude for the value of $\epsilon_{M} \sim 10^{-2}$ TK$^{-1}$ for yttrium iron garnet (Y$_3$Fe$_5$O$_{12}$). 
\end{abstract}


%
%

\maketitle
\section{Introduction}
\label{sect:introduction}

The spin Seebeck effect consists in a spin or magnetization current generated by a temperature gradient across a ferromagnetic material. While the possibility of such effect is expected in analogy to the well known thermoelectric effects, its experimental verification has been the subject of several attempts with many different configurations and setups \cite{Gravier-2006, Uchida-2008, Uchida-2010}. The most interesting and promising combination between materials and configurations is the longitudinal spin Seebeck effect (LSSE) found in ferromagnetic insulators \cite{Uchida-2010}. Even if the spin Seebeck effect has been revealed in different magnetic materials \cite{Uchida-2010b}, the most studied one is the yttrium iron garnet (Y$_3$Fe$_5$O$_{12}$, YIG) which exhibits large effects \cite{Siegel-2014, Uchida-2014}. While one associates the thermoelectric effects in metals to the transport properties of electrons, the effect present in a magnetic insulator is thought to be related to the non equilibrium spin waves (magnons) carrying both magnetization and heat when the sample is in a temperature gradient \cite{Adachi-2013}. Much of the experimental efforts have been devoted to the detection of the spin currents generated by the spin Seebeck effect. The most successful method is the detection of the transverse voltage induced in a thin paramagnetic platinum layer placed at the YIG surface \cite{Uchida-2013}. The principle of the detection of the spin current is the inverse spin Hall effect (ISHE) in metals. Due to the spin orbit coupling of conduction electrons, a spin polarized electron is scattered perpendicularly to both the direction of the motion and the direction of the spin, by an angle $\pm \theta_{SH}$, depending on its up or down spin state. One may then detect the presence of a spin current through the detection of an electric potential in the direction of the scattering. As the ISHE Pt sensor can be placed both parallel as well as perpendicular to the YIG temperature gradient (and therefore to the main YIG spin current), in the literature one finds a distinction between: i) the longitudinal effect, when the Pt senses a spin current parallel to the YIG temperature gradient and ii) the transverse effect when the Pt senses a spin current perpendicular to the YIG temperature gradient. Both longitudinal and transverse spin Seebeck experiments reveal a linear dependence of the ISHE voltage as a function of the temperature gradient. This means that the spin current of magnons in the YIG is  converted into a spin current carried by electrons inside the platinum by a coupling between the localized magnetic moment of the YIG and the conduction electrons of Pt \cite{Uchida-2013}.

From the thermodynamics viewpoint, the spin Seebeck phenomena can be analyzed by making an analogy (or an extension) of the thermoelectric effects by considering the spin at the place of (or in addition to) the charge \cite{Bauer-2012}. However one of the key points of the detection longitudinal spin Seebeck effect is the passage of the spin current through different materials (YIG insulators and Pt metal) which calls for a better understanding of the evolution of the currents and the driving forces in different materials. In the present paper we investigate the issue from the viewpoint of the non equilibrium thermodynamics of fluxes and forces to describe spin and heat transport. For what concerns the spin degree of freedom the theory can be equivalently developed in terms of spin current or magnetization current and we choose the second one as in Ref.\cite{Johnson-1987}. After the pioneering paper by Johnson and Silsbee \cite{Johnson-1987}, it has became clear that the main difference with respect to the classical theories of the thermoelectric effects is that the magnetization current density $j_M$ is not continuous and therefore one of the crucial points, also in the comparison with experiments, is to understand the profile of $j_M$ through the different layers. To this aim one has first to provide a proper definition for the thermodynamic driving force associated with the magnetization current in YIG, an aspect which is not enough clear in the literature. Starting from the thermodynamic approach of Johnson and Silsbee \cite{Johnson-1987} and stating the continuity equation for the magnetization, we are able to identify the thermodynamic driving force of the magnetization current as the gradient of an effective field $\nabla H^*$. This definition permits to derive the profiles of the magnetization current along different media. In the paper we apply the theory to the geometry of the longitudinal spin Seebeck effect with YIG and Pt. By focusing on the specific geometry with one YIG layer and one Pt layer, we obtain the optimal conditions for generating large magnetization currents. Furthermore, by using the theory one can possibly have access to the spin Seebeck coefficient $\epsilon_M$ relating, in correspondence to its electric analogue, the gradient of the potential of the magnetization current, $\mu_0\nabla H^*$ to the temperature gradient $\nabla T$. By using results of recent experiments and literature data \cite{Uchida-2010, Sola-2015}, we are able to obtain an order of magnitude for the spin Seebeck coefficient $\epsilon_{M}$ of the YIG as $\epsilon_{M} \sim 10^{-2}$ TK$^{-1}$ that can be compared with the theories of magnon diffusion.

\begin{figure}[htb]
	\begin{centering}
	\includegraphics[width=1\textwidth]{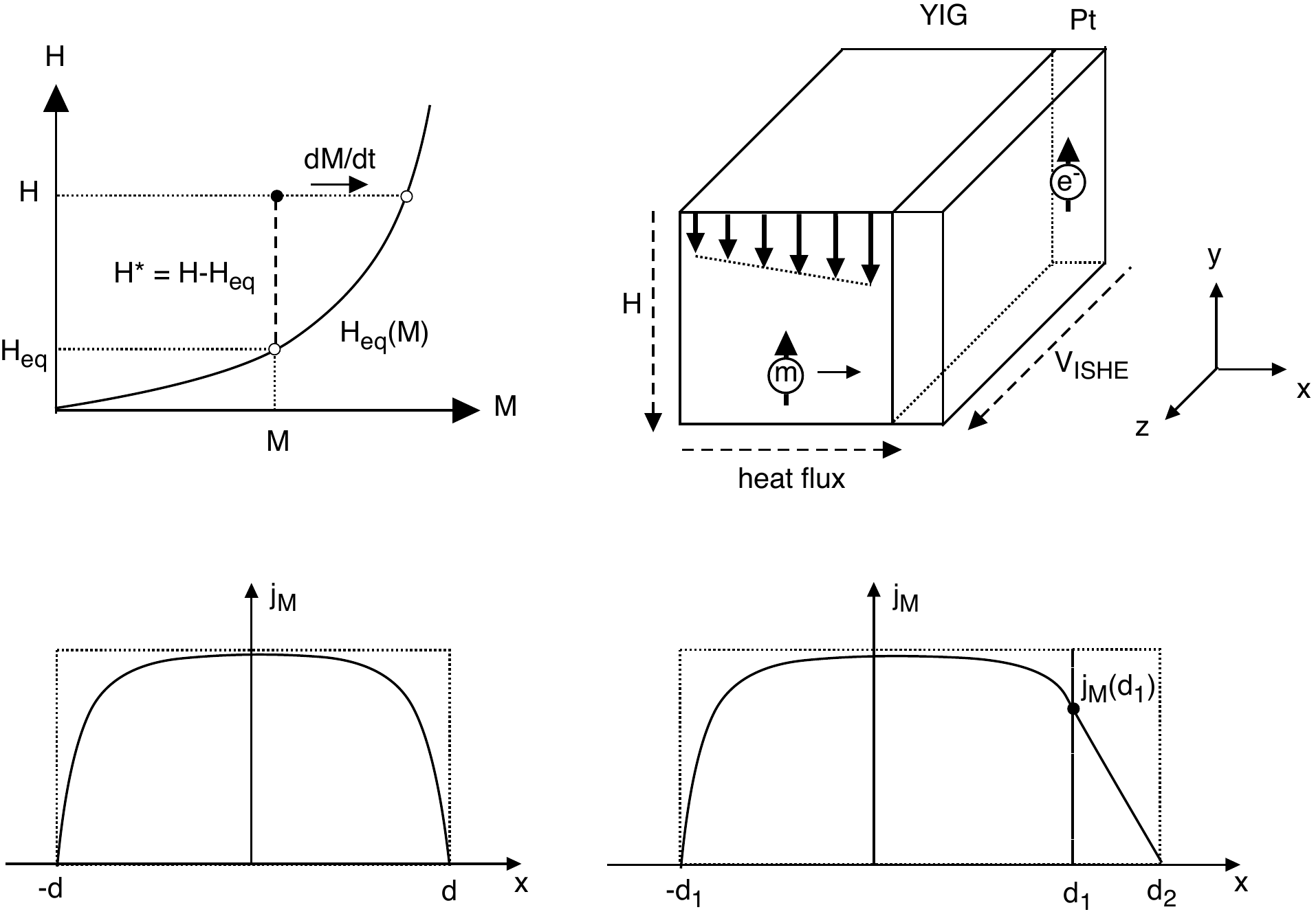}
	\caption{Top left: non equilibrium condition for the local magnetization. Top right: geometry of the longitudinal spin Seebeck effect. Bottom left: boundary conditions for the spin Seebeck effect in a single active material. Bottom right: boundary conditions for the longitudinal spin Seebeck with an active material and a sensing layer.}
	\label{FIG:LSSE}
	\end{centering}
\end{figure}

\section{Thermodynamic theory of magnetization transport and the spin Seebeck coefficient}
\label{sect:Thermodynamic}

To define the thermodynamic driving force of the magnetization current we first state a continuity equation for the non conserved magnetization. We will use throughout the paper the international system of units. We consider here the case of scalar magnetization and we take the following continuity equation

\beq
\frac{\partial M}{\partial t}+\nabla \cdot j_M = \frac{ H - H_{eq}}{\tau_M}
\label{EQ:cont}
\eeq

\noindent where $M$ is the magnetization (the volume density of magnetic moment, measured in Am$^{-1}$), $j_M$ is the magnetization current density, $H$ is the magnetic field, $H_{eq}(M)$ the magnetic equation of state at equilibrium and $\tau_M$ is a relaxation time associated with the damping process. The equation means that every time the field $H$ is not equal to the equilibrium value $H_{eq}$, either the local magnetization changes in time or a magnetization current is established (see Fig.1 top left). The term on the right hand side represents then the sources and sinks for the non conserved magnetization. The non equilibrium thermodynamics of fluxes and forces can be well developed in the enthalpy $u_e = u - \mu_0HM$ representation, with $u$ the internal energy, where the independent variables are the magnetic field $H$ and the entropy $s$. By expressing the change in enthalpy in the general case of a magnetic field different from the equilibrium value $H_{eq}(M)$, we arrive at the following equation

\beq
du_e = Tds -\mu_0 MdH - \mu_0 \left( H - H_{eq}\right)dM
\eeq

\noindent By using the previous equation to define the currents of the intensive variables (entropy, energy and magnetization) one finds that the thermodynamic driving force for the magnetization current is the gradient of an effective field defined as 

\beq
H^* = H - H_{eq}
\eeq

\noindent The detailed derivation follows the classical route \cite{Callen-1985} and will be presented elsewhere \cite{Basso-2015}. Here we employ this main result to state, in perfect analogy with the theory of thermoelectric effects \cite{Callen-1985}, the equations relating the magnetization current and the heat current to the associated forces. We limit to currents and forces in one dimension ($\nabla = \partial/\partial x$). The equations are 

\begin{align}
\label{EQ:M_curr}
j_M & = \sigma_M \, \mu_0 \nabla H^* - \epsilon_M \sigma_M \, \nabla T \\
j_q & = \epsilon_M \sigma_M T \mu_0\nabla H^* - (\kappa_M + \epsilon_M^2 \sigma_M T) \nabla T
\label{EQ:q_curr}
\end{align}

\noindent where $\sigma_M$ is the spin conductivity, $\epsilon_M$ is the spin Seebeck coefficient, $j_q$ is the heat current density and $\kappa_M$ is the spin thermal conductivity. Since the magnetization is not conserved, the magnetization current is not continuous and we  always have to specify the geometrical conditions we want to investigate. As we are interested to non-equilibrium stationary states we always ask the condition $\partial M/\partial t=0$ to be true, so, the continuity equation (Eq.(\ref{EQ:cont})) becomes

\beq
\tau_M \nabla j_{M} = H^*
\label{EQ:cont_jM}
\eeq

\noindent If we disregard for the moment the heat current, the solution of the magnetization current problem will correspond to find solutions to the system composed by Eq.(\ref{EQ:M_curr}) and Eq.(\ref{EQ:cont_jM}) by posing the appropriate boundary conditions. By putting them together and assuming that the second term of the right hand side of Eq.(\ref{EQ:M_curr}) does not depend on $x$, we obtain a differential equation for the driving potential

\beq
l_M^2 \nabla^2 H^* = H^*
\label{EQ:H*}
\eeq

\noindent where $l_M$ is 

\beq
l_M = \sqrt{\mu_0 \sigma_M\tau_M }
\label{EQ:l_M}
\eeq

\noindent The differential equation has general solutions in the form of exponentials $\exp(-x/l_M)$ and the specific solution only depends on the boundary conditions. The length $l_M$ can then be interpreted as a diffusion length for the spin transport process in a specific material. If the theory is applied to electric conductors, the effective field $H^*$ turns out to be the same as the spin chemical potential introduced in \cite{Johnson-1987, Valet-1993}. The interest in the present treatment is to apply to magnetic insulators and to the junctions between magnetic insulators and normal conductors. 

\begin{figure}[htb]
	\begin{centering}
	\includegraphics[width=1\textwidth]{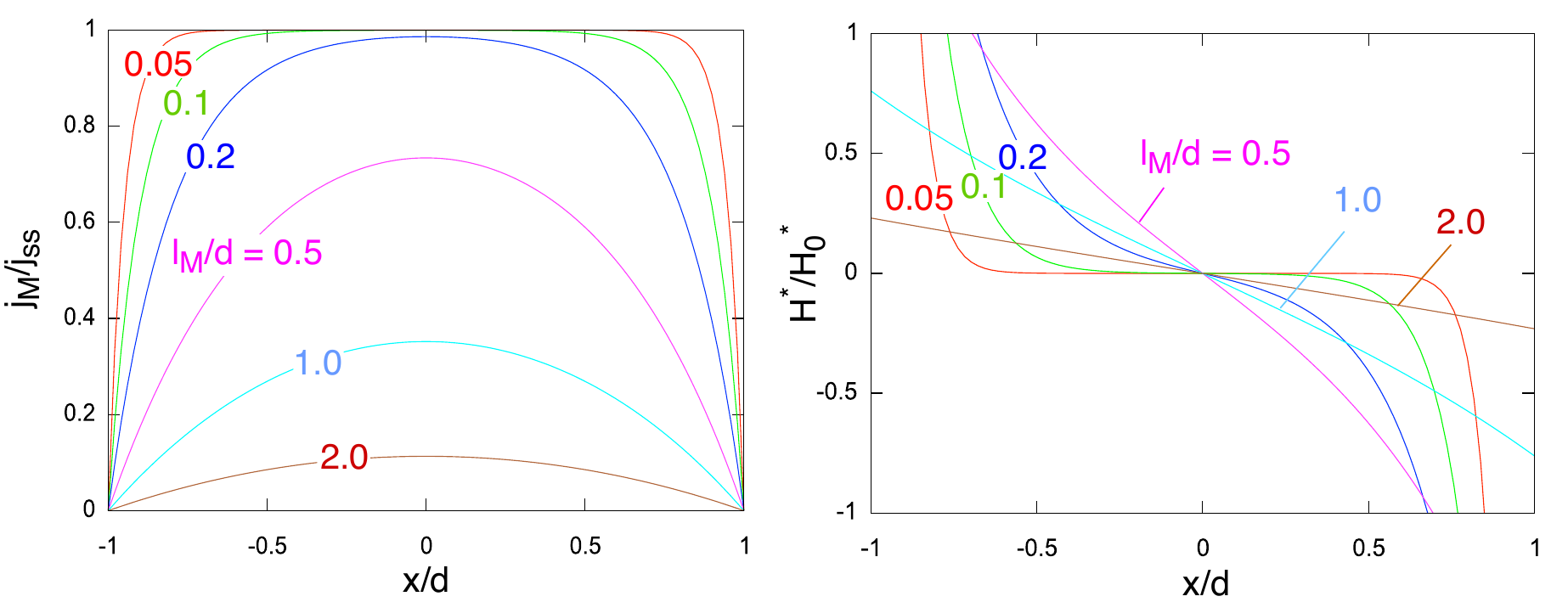}
	\caption{Longitudinal spin Seebeck effect in a single material. Left: magnetization current profiles from Eq.(\ref{EQ:1mat_j}). Right: effective field profile $H^*(x)$ from Eq.(\ref{EQ:1mat_H}) with $H_0^{*}=j_{ss}/(l_M/\tau_M)$}
	\label{FIG:1mat}
	\end{centering}
\end{figure}

\begin{figure}[htb]
	\begin{centering}
	\includegraphics[width=1\textwidth]{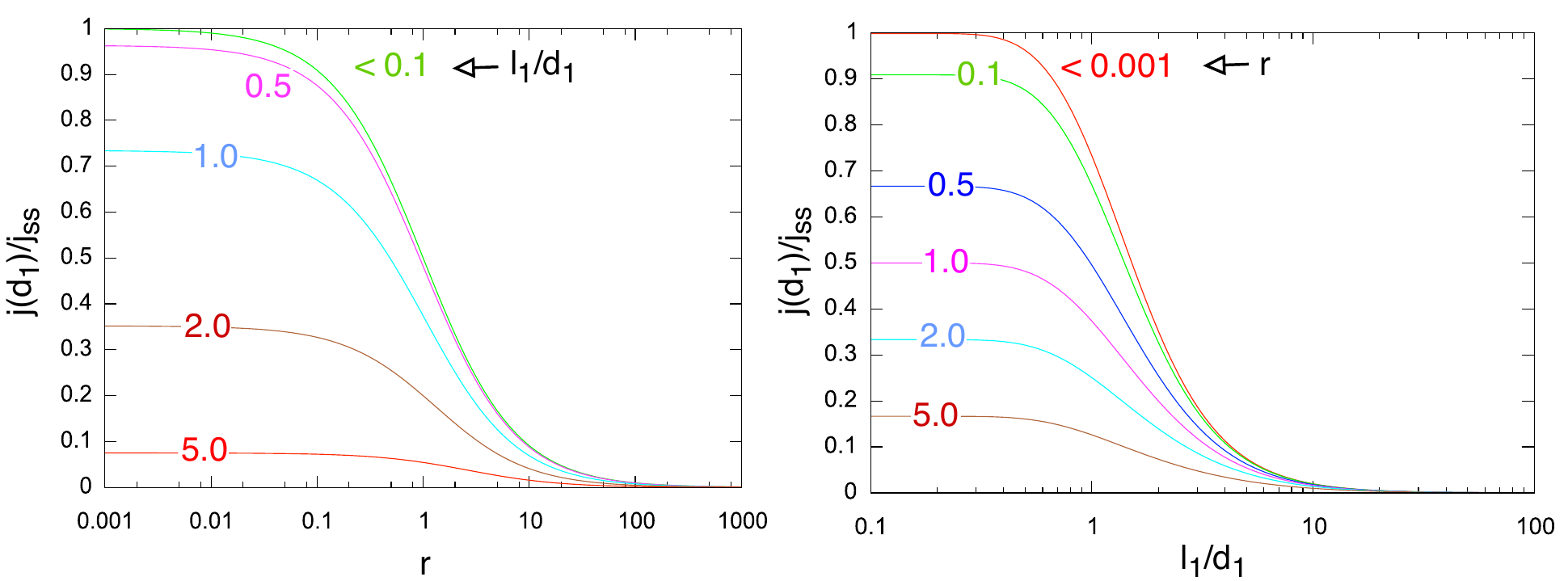}
	\caption{Longitudinal spin Seebeck effect at the junction between an insulator (1) and a metal (2). Both panels show the magnetization current at the interface $j(d_1)$ from Eq.(\ref{EQ:inj_cond}) in the limit $l_2 << t_2$. $j(d_1)$ is normalized to the maximum value $j_{ss}$ and is shown as a function of the ratio $l_1/d_1$ of the diffusion length $l_1$ over the half thickness $d_1$ of the insulator and of the ratio $r = (l_1/\tau_1)/(l_2/\tau_2)$. }
	\label{FIG:2mat}
	\end{centering}
\end{figure}

\section{Application of the theory to the geometry of LSSE}
\label{sect:Application}

We first consider a magnetic insulator of length $t=2d$ with a magnetization current along $x$ due to the spin Seebeck effect \cite{Uchida-2010,Sola-2015} (see Fig.1 bottom left). We consider the insulator subjected to a constant temperature gradient, therefore the magnetization current is given by Eq.(\ref{EQ:M_curr}). We define the spin Seebeck current as 

\beq
j_{ss}=- \epsilon_M \sigma_M \, \frac{\partial T}{\partial x}  
\eeq

\noindent giving then, from Eq.(\ref{EQ:M_curr}),

\beq
j_M = \mu_0 \sigma_M \, \frac{\partial H^*}{\partial x} + j_{ss}
\label{EQ:cont_jM1}
\eeq

\noindent We now have to solve Eq.(\ref{EQ:H*}) with boundary conditions $j_M(-d)=j_M(d)=0$ imposed through Eq.(\ref{EQ:cont_jM1}). The solution for the effective field is

\beq
H^*(x) = - \frac{j_{ss}}{(l_M/\tau_M)}\frac{\sinh(x/l_M)}{\cosh(d/l_M)}
\label{EQ:1mat_H}
\eeq

\noindent and, from Eq.(\ref{EQ:cont_jM1}), the magnetization current is

\beq
j_M(x) = j_{ss} \left( 1- \frac{\cosh(x/l_M)}{\cosh(d/l_M)} \right)
\label{EQ:1mat_j}
\eeq

\noindent Fig.\ref{FIG:1mat} shows the profiles of the magnetization current and the effective field along the material for different values of $l_M$. From the pictures one clearly sees how the spin accumulation close to the boundaries generates, as a reaction, an effective field which counteracts the spin Seebeck effect in order to let the current to go to zero at the interface. 

The situation changes if we consider instead the geometry of the spin Seebeck experiments of Refs.\cite{Uchida-2010, Sola-2015} (see Fig.1 bottom right) in which we have the longitudinal diffusion of the spin current from the YIG to the Pt. Here the boundary conditions have to be set to allow the possibility for the magnetization current to proceeds between different media. The quality of the YIG/Pt interface in known to play an important role \cite{Qiu-2015}. In the model it can be taken into account by introducing a third effective layer between the two with degraded properties. For simplicity we consider here an ideal interface between YIG and Pt. We then have media 1 (the YIG ferrimagnetic insulator) of thickness $t_1=2d_1$ from $x=-d_1$ to $x=d_1$ and media 2 (the Pt non magnetic conductor) of thickness $t_2 = d_2-d_1$ from $x=d_1$ to $x=d_1$. For simplicity we drop all the $M$ subscripts and we replace them with the index (1 or 2) of the corresponding media. In YIG (1), with the boundary condition $j_1(-d_1) = 0$, we have

\beq
j_{1}(x) = j_{ss}\left( 1-\frac{\cosh(x/l_1)}{\cosh(d_1/l_1)} \right) + j_1({d_1}) \frac{\sinh((x+d_1)/l_1)}{\sinh(t_1/l_1)}
\label{EQ:inj_j1}
\eeq

\noindent and

\beq
H^*_1(x) = - \frac{j_{ss}}{ (l_1/\tau_1)} \frac{\sinh(x/l_1)}{\cosh(d_1/l_1)} + \frac{j_1(d_1)}{ (l_1/\tau_1)} \frac{\cosh((x+d_1)/l_1)}{\sinh(t_1/l_1)}
\label{EQ:inj_H1}
\eeq

\noindent while in Pt (2) we have

\beq
j_{2}(x) = j_2({d_1}) \frac{\sinh((d_2-x)/l_2)}{\sinh((d_2-d_1)/l_2)}
\label{EQ:inj_j2}
\eeq

\noindent and

\beq
H^*_{2}(x) = - \frac{j_2(d_1)}{(l_2/\tau_2)} \,\, \frac{\cosh((d_2-x)/l_2)}{\sinh((d_2-d_1)/l_2)}
\label{EQ:inj_H2}
\eeq

\noindent By setting the boundary condition at the interface between the two media $j_1(d_1) = j_2(d_1)$ and $H^*_{1}(d_1)=H^*_{2}(d_1)$ we find the value of the current at the interface

\beq
j(d_1) = j_{ss} \frac{(l_2/\tau_2) \tanh(d_1/l_1)}{(l_2/\tau_2) \coth(t_1/l_1) + (l_1/\tau_1)  \coth(t_2/l_2)}
\label{EQ:inj_cond}
\eeq

\noindent The magnetization current and the effective field are then obtained by the substitution of the current at the interface (Eq.\ref{EQ:inj_cond}) into Eqs.(\ref{EQ:inj_j1}) and (\ref{EQ:inj_j2}) for the current and Eqs.(\ref{EQ:inj_H1}) and (\ref{EQ:inj_H2}) for the field. Eq.(\ref{EQ:inj_cond}) permits to obtain the optimal conditions for generating large magnetization currents across the interface. Some insight into the effectiveness of the spin injection from YIG into Pt can be gained by taking the limit $l_2 << t_2$ in which the magnetization current of media (2) (Eq.(\ref{EQ:inj_j2})) becomes an exponential decay $j_2(x) = j(d_1) \exp((d_1-x)/l_2)$. Fig.\ref{FIG:2mat} shows the magnetization current at the interface $j(d_1)$ normalized to the maximum value $j_{ss}$ as a function of the ratio $l_1/d_1$ of the diffusion length $l_1$ over the half thickness $d_1$ of the insulator and of the ratio $r = (l_1/\tau_1)/(l_2/\tau_2)$.  From Fig.\ref{FIG:2mat} we have that the injection into the media (2) is effective only in the conditions of $l_1<d_1$ and $(l_1/\tau_1)<(l_2/\tau_2)$. These conditions depend on the intrinsic properties of the media and on the thickness of the YIG.

\section{Comparison with experiments}
\label{sect:Comparison}

Recent experiments performed on a system composed by a YIG layer of thickness $t_{1} = 4 \, \mu$m and a platinum layer of thickness $t_{2} = 10$ nm and width $w_{Pt} = 6$ mm, have revealed a LSSE coefficient $E_z/(\nabla_x T)= 2.8\cdot10^{-7}$ VK$^{-1}$ \cite{Sola-2015}. The system is described in Fig.1 top right.  The heat flux is applied along $x$, the magnetic field is along $-y$, the ISHE voltage is measured along $z$ and $E_z = V_{ISHE}/w_{Pt}$. The magnetization current is along $x$ and the direction of the magnetic moment of the diffusing species (magnons or electrons) is along $y$. Before comparing the theory results to the LSSE experiments we have to evaluate spin current at the interface from the ISHE voltage. The ISHE is described by the spin Hall angle giving the ratio of the component of the current deflected by the spin orbit coupling over the imposed current $\theta_{SH} = (j_{ez}/e)/(j_{Mx}/\mu_B)$, where $\mu_B$ is the Bohr magneton and $e$ is the elementary charge. In the LSSE experiment the measurement is performed in open circuit, so one uses $j_{ez} = \sigma_{Pt} E_z$, where $\sigma_{Pt} = 6.4 \cdot 10^6 \, \Omega^{-1}$m$^{-1}$ is the electrical conductivity of Pt. The average magnetization current $<j_{Mx}>$ diffusing into the Pt is then given by

\beq
<j_{Mx}>=\frac{\mu_B}{e}\frac{1}{\theta_{SH}}j_{ez} = \frac{\mu_B}{e}\frac{\sigma_{Pt}}{\theta_{SH}}E_z
\eeq

\noindent From Ref.\cite{Wang-2014} we take $\theta_{SH} =0.1$ for Pt and obtain an average current $<j_{Mx}>/(-\nabla_x T) \simeq 1.0 \cdot10^{-3}$ As$^{-1}$K$^{-1}$m. From Ref.\cite{Wang-2014} we also obtain a value for the diffusion length of Pt as $l_{2}=7.3$ nm. The spin conductivity of Pt can be estimated by assuming that in a normal metal the scattering acts independently of the spin \cite{Valet-1993}. Then, by converting the conductivity of Pt into the conductivity of the magnetization current, we obtain $\mu_0\sigma_2=2.6 \cdot 10^{-8}$ m$^{2}$s$^{-1}$ and consequently a time constant $\tau_2 = l_2^2/(\mu_0\sigma_2) \simeq 2 \cdot 10^{-9}$ s. By using these numbers into Eq.(\ref{EQ:inj_j2}) one finds that the profile of the magnetization current is, at a good approximation, a linear decay from the interface to the border. Therefore we take for simplicity that the current at the interface is twice the average estimated value, then $j(d_1)/(-\nabla_x T) \simeq  2 \cdot10^{-3}$ As$^{-1}$K$^{-1}$m. The most critical part is now the estimation of the diffusion length inside YIG, $l_1$, the time constant, $\tau_1$, and the conductivity, $\sigma_1$, which are not exactly known. For the transverse experiment (in which current and magnetization are parallel) estimates range from micron to millimeter \cite{Uchida-2012, Giles-2014arXiv, Cornelissen-2015arXiv}. For the longitudinal effect (in which current and magnetization are perpendicular) it is believed to be much less (i.e. $< 1 \mu$m) \cite{Ramos-2015arXiv}. We attempt here an order of magnitude estimate based on the results obtained in the previous section and on literature experiments. From Ref.\cite{Uchida-2014} the spin Seebeck coefficient measured on 1 mm of YIG results to be larger $E_x/(\nabla_x T) \simeq 4\cdot10^{-7}$ VK$^{-1}$ than the one of the 4 $\mu$m sample \cite{Sola-2015}, but of the same order of magnitude. Therefore, from Fig.3, we can guess that the $l_1$ should be of the same order of magnitude of the thinner sample (4 $\mu$m) in order to allow for an efficient injection in both cases. We set $l_1 = 1 \mu$m and, to have a reasonable injection (50\%, i.e. $j(d_1)= 0.5 \, j_{ss}$, see Fig.3), we also set $r=1$, i.e. $l_1/\tau_1 = l_2/\tau_2$. By using the resulting value $\mu_0\sigma_1 \sim 4 \cdot 10^{-7}$ m$^{2}$s$^{-1}$, we can finally obtain an order of magnitude for the spin Seebeck coefficient of Eq.(\ref{EQ:M_curr})

\beq
\epsilon_{1} = \frac{1}{\sigma_{1}}\left( \frac{j_{ss}}{-\nabla T}\right)
\eeq

\noindent as $\epsilon_{1} \sim 10^{-2}$ TK$^{-1}$. This value should be correlated with the theories of the magnon diffusion which will be the subject of future work. To give an order of magnitude, as for the thermoelectric effects the Seebeck coefficient is compared to the classical value $\epsilon = -k_B/e \simeq - 86 \cdot 10^{-6}$ VK$^{-1}$, the value found here can be compared with the ratio $k_B/\mu_B \simeq 1.49 $ TK$^{-1}$. 

\section{Conclusions}
\label{sect:acks}

In this paper we have employed the non equilibrium thermodynamics of fluxes and forces to describe the space profiles of the magnetization current for the geometry of the longitudinal spin Seebeck experiment. The theory permits to derive the conditions to have an efficient longitudinal spin injection from YIG to Pt. By the theory we are able to define the spin Seebeck coefficient $\epsilon_M$. Even if many material dependent parameters are not available yet, a rough order-of-magnitude estimate gives $\epsilon_{M} \sim 10^{-2}$ TK$^{-1}$ for the YIG. More quantitative estimations are expected as soon as LSSE measurements with different YIG thicknesses will be performed. The theory used in the paper, which describes the spin currents in the phenomenological framework of thermodynamics, has possible interesting applications also to the current profiles seen in the transverse spin Seebeck experiments in which the currents in Pt diffuse perpendicularly to the main magnetization current of the YIG. To this aim the vector extension of the present theory is seen as a possible and future development. 

\section{Acknowledgments}
\label{sect:acks}

This work has been carried out within the Joint Research Project EXL04 (SpinCal), funded by the European Metrology Research Programme. The EMRP is jointly funded by the EMRP participating countries within EURAMET and the European Union.

%
\bibliography{00_SS}
\bibliographystyle{plain}



\end{document}